\documentclass[letter]{aa}  

\usepackage{graphicx}
\usepackage{txfonts}
\begin{document}

\titlerunning{Pluto's atmosphere in 2019}
\authorrunning{Arimatsu et al.}

   \title{Evidence for a rapid decrease of Pluto's atmospheric pressure revealed by a stellar occultation in 2019}

   \author{K. Arimatsu\inst{1}
   		%I'm DORAEMON!!!
          \and
          G. L. Hashimoto\inst{2}
          \and
          M. Kagitani\inst{3}
          \and
          T. Sakanoi\inst{3}
          \and
          Y. Kasaba\inst{3}
          \and
          R. Ohsawa\inst{4}
          \and
          S. Urakawa\inst{5}
          }

   \institute{Astronomical Observatory, Graduate School of Science,  Kyoto University Kitashirakawa-oiwake-cho, Sakyo-ku, Kyoto 606-8502, Japan\\
              \email{arimatsu@kwasan.kyoto-u.ac.jp}
          	\and
             	Department of Earth Science, Okayama University, 1-1-1 Kita-ku Tsushimanaka, Okayama 700-8530, Japan
             	\and
		 Planetary Plasma and Atmospheric Research Center, Graduate School of Science, Tohoku University, 6-3 Aramaki -aza-aoba, Aoba-ku, Sendai 980-8578, Japan
		\and
		Institute of Astronomy, Graduate School of Science, The University of Tokyo, 2-21-1 Osawa, Mitaka, Tokyo 181-0015, Japan
		\and
		Japan Spaceguard Association, Bisei Spaceguard Center 1716-3 Okura, Bisei, Ibara, Okayama 714-1411, Japan	
             }

    \date{}

\abstract
{
We report observations of a stellar occultation by Pluto on 2019 July 17.
A single-chord high-speed (time resolution = 2\,s) photometry dataset was obtained with a CMOS camera mounted on the Tohoku University 60 cm telescope (Haleakala, Hawaii). 
The occultation light curve is satisfactorily fitted to an existing Pluto's atmospheric model. 
We find the lowest pressure value at a reference radius of $r = 1215~{\rm km}$ among those reported after 2012, indicating a possible rapid (approximately  $21^{+4}_{-5} \%$ of the previous value) pressure drop between 2016 (the latest reported estimate) and 2019. 
However, this drop is detected at a $2.4\sigma$ level only and still requires confirmation from future observations. 
If real, this trend is opposite to the monotonic increase of Pluto's atmospheric pressure reported by previous studies.
The observed decrease trend is possibly caused by ongoing ${\rm N_2}$ condensation processes in the Sputnik Planitia glacier associated with an orbitally driven decline of solar insolation, as predicted by previous theoretical models.
However, the observed amplitude of the pressure decrease is larger than the model predictions.
}

\keywords{Kuiper belt objects: individual (Pluto) --- occultations --- planets and satellites: atmospheres --- planets and satellites: physical evolution}

\maketitle	

\section{Introduction} \label{sec:intro}

The existence of its substantial atmosphere primarily consisting of ${\rm N_2}$ with trace amounts of ${\rm CH_4}$, ${\rm CO}$, ${\rm HCN}$, and other species makes Pluto unique among known trans-Neptunian objects (TNOs) 
since no clear observational evidence for an atmosphere with a surface pressure greater than $\sim 10 \ {\rm nbar}$ is found on the other bodies \citep{Sicardy11,Ortiz12,Arimatsu19b}.
After its confirmation in 1988 \citep{Hubbard88,Elliot89}, stellar occultation observations have been a powerful tool to explore Pluto's atmosphere; clear refraction features in the light curve of an occulted star probes the number density, temperature, and pressure profiles.
Pluto's stellar occultations especially provide unique opportunities to monitor seasonal evolutions of the atmospheric pressure, which was discovered by \citet{Sicardy03}.
The pressure evolution is caused by surface sublimation-condensation processes of Pluto's ${\rm N2}$ and the other volatiles, which are thought to account for the large diversity of the surface structures revealed by the {\it New Horizons} spacecraft (hereafter NH; \citealt{Stern15}).
Revealing the pressure evolution through stellar occultations is thus a key not only to obtaining basic physical properties of the atmosphere but also to understanding the geology of this unique TNO.

Although Pluto has receded from the Sun since 1989, its observed atmospheric pressure shows a monotonic increase (e.g., \citealt{Sicardy03, Young13, Olkin15, Dias-Oliveira15, Sicardy16, Meza19}). 
The observed increase trend has been modeled \citep{Bertrand16, Forget17, Meza19} based on surface topographic features and volatile ice distributions observed by NH \citep{Moore16}.
According to the models, the surface ${\rm N_2}$ ice glacier covering Sputnik Planitia, which is the largest confirmed basin, plays an important role in the seasonal pressure evolution.
Model predictions indicate that  
the pressure was increasing and reached its maximum value in $\sim 2015$.
These models predict the decrease of pressure after 2015-2020 since the insolation in Sputnik Planitia decreased. 
Further continuous observations are required to test the model predictions.
Unfortunately, Pluto has been receded away from the Galactic plane,
and the probability of occultation events has decreased since $\sim 2015$.
Observations of stellar occultations thus become rare opportunities to examine the current state of Pluto's atmosphere.

In this Letter, we report the observations of a stellar occultation by Pluto on  2019 July 17, approximately three years after the latest reported stellar occultation event (2016 July 19; \citealt{Meza19}). 
Photometric data of the occultation have been obtained 
with a CMOS camera mounted on the Tohoku University 60 cm telescope (Haleakala, Hawaii).
Section 2 presents the outline of the stellar occultation observation
and the data reduction method to derive the light curve.
The results of the observation and the pressure estimate are given in Section 3.
Discussions based on the obtained results are provided in Section 4.
Finally, we summarize the results and discussions in Section 5.

\section{Observation And Data Reduction} \label{sec:obs}

\subsection{Observation}

The occultation of a star UCAC5 340-173206 (Gaia DR2 catalog source ID: 6772059623498733952, ICRS position at epoch J2000: $\alpha =  19^{\rm h}~33^{\rm m}~26^{\rm s}.013$, $\delta = -22\degr~07'~58.42"$, Gaia Gmag = 13.0; \citealt{Gaia18}) on 2019 July 17 was originally predicted by ERC Lucky Star project\footnote{\url{http://lesia.obspm.fr/lucky-star/occ.php?p=13163}}. 
The project predicts stellar occultations by Pluto using {\it Numerical Integration of the Motion of an Asteroid} (NIMA) version 8 orbital elements \citep{Desmars19}. 
The prediction confirmed that Pluto's shadow swept over the major islands of Hawaii at a geocentric velocity $v_{\rm rel} = 24.18~{\rm km~s^{-1}}$. 
The geocentric distance of Pluto during the occultation corresponds to $D = 32.83~{\rm au}$.

On 2019 July 17, we observed UCAC5 340-173206 at the Tohoku University Haleakala Observatory (latitude: $20\degr$ $42'$ $30"$ N, longitude: $156\degr 15'  30"$ W, altitude: 3040~m), Hawaii. 
At Haleakala, photometric data were obtained using a ZWO Co., Ltd. ASI178MM CMOS camera mounted on the Coude focus of the Tohoku University 60 cm Cassegrain / Coude (T60) telescope.
The field of view (FoV) for the CMOS camera is $120\arcsec \times 80\arcsec$  with an angular pixel scale of 0.\arcsec038. 
The exposure time was 2~s for each frame corresponding to the frame rate of $\sim 0.5~{\rm Hz}$.
No filter was used in the present observation.
During the observation, a dataset consisting of 1305 image frames were obtained in a 43.5-minute window centered on the predicted central time of the occultation (12:42 on 2019 July 17 UTC) with an imaging capture software {\it Firecapture} version 2.4.

\subsection{Data reduction}

Aperture photometry for the occulted star (plus Pluto and Charon) is performed using the image data after the bias and flat-field corrections and the sky background subtraction. 
Due to possible rapid changes in atmospheric conditions, 
the obtained flux values for individual data points can suffer time-dependent flux fluctuations.
To correct the possible flux fluctuations and calibrate the flux value of the target star, 
we carry out differential photometry with three reference stars (Gaia DR2 G-band magnitude range of $11 - 14$) that are detected simultaneously. 
The calibrated light curve is then normalized to the unocculted flux. 

We also obtained a Pluto $+$ Charon flux relative to the total flux (including the occulted star)
with resolved images obtained approximately 40 minutes before the occultation event.
To obtain an accurate Pluto $+$ Charon flux, we subtract a radial brightness profile of UCAC5 340-173206 located close to Pluto and Charon in the original images based on a digital coronagraphic method developed by \citet{Assafin09}.
The derived relative flux ratio is $\sim 0.241$.
We should note that, however, the imperfect flat-field correction, the residual of the stellar radial brightness profile, the airmass variation (1.5 during calibration versus 1.7 during the occultation), and Pluto's rotational albedo variation can cause a systematic uncertainty of the flux ratio.
We thus use this flux ratio only as a reference value for the light curve fit (see Section~3.1).

\section{Results}
\subsection{Light curve fitting}

Figure~1 shows the light curve of the target star after the calibration and the normalization.
The observed duration of the occultation is roughly $\sim 50$~s.
No significant asymmetric profile or spike-like feature was detected from the light curve.
To obtain the pressure of the atmosphere, 
 synthetic light curves as a function of the pressure at a radius of $r = 1215~{\rm km}$, which has been referred to be a reference radius by previous studies \citep{Yelle97, Sicardy16, Meza19}, $p_{1215}$, the distance of closest approach of Haleakala Observatory to Pluto's shadow center, $\rho$ the central time of the occultation, and the Pluto $+$ Charon flux relative to the total flux,
are generated using a ray-tracing technique described by \citet{Sicardy99}.
For Pluto's atmospheric refraction model, we consider atmospheric profiles provided by \citet{Dias-Oliveira15}
with assumptions that Pluto has a pure ${\rm N_2}$, spherically symmetric and transparent (haze-free) atmosphere with a time-independent thermal structure derived from the observed light curves.
In the present model, the contribution from the near-limb (primary) image and far-limb (secondary) images are considered with a focusing factor  
(the ratio of the observer’s distance to the shadow center and the closest approach of the corresponding ray in Pluto’s atmosphere assuming a circular limb curvature of Pluto) provided by \citet{Sicardy99}.
The validity of these assumptions is discussed by \citet{Meza19} with the NH results.
According to \citet{Meza19}, the fixed temperature profile appears to be close to the results derived from NH and is thus useful for at least estimating the relative pressure changes.
The physical parameters used for the fit are presented in Table~1.
We adopted these parameters following \citet{Dias-Oliveira15} and \citet{Meza19} to avoid systematic uncertainties caused by using different values.

The synthetic light curve is fitted to the observed data points (77 data points shown in Figure~1 are used for the fit) by minimizing $\chi^2$; 
\begin{equation}
\label{eq_chi}
\chi^2 = \sum_i \frac{(\phi_{i, {\rm obs}} - \phi_{i, {\rm syn}})^2}{\sigma_i^2},
\end{equation}
where $\phi_{i, {\rm obs}}$, $\phi_{i, {\rm syn}}$, and $\sigma_i$ are the observed and synthetic stellar fluxes, and the $1\sigma$ error at a data point $i$, respectively. 
Figure~2 shows the $\chi^2$ map for the fit to the light curve as a function of $p_{1215}$ and $\rho$.
The $\chi^2$ value for the best fit is 61.9 with 73 degrees of freedom (corresponding to 77 data points - 4 free parameters), indicating a satisfactory fit with a reduced $\chi^2$ of $\sim 0.848$.
The best-fit parameters are shown in Table~1, and the best-fit synthetic data points are overlaid with the observed data points in Figure~1. 
According to the best-fit model, observed stellar rays come from above $r \sim 1214~{\rm km}$, which is comparable to the adopted reference radius.
The obtained reference radius pressure is $p_{1215}  = 5.20^{+0.28}_{-0.19}~{\rm \mu bar}$ ($1\sigma$ level error bars, see Figure~2).
Assuming the lower atmosphere temperature profile provided by \citet{Dias-Oliveira15}, this value corresponds to a surface ($r = 1187~{\rm km}$) pressure of $p_{\rm surf} = 9.56^{+0.52}_{-0.34}~{\rm \mu bar}$.

The best-fit $\rho$ is derived to be $\rho = 1008.0^{+7.8}_{-7.2}~{\rm km}$. 
The central time of the occultation derived from the fit is 12:41:$52.02\pm 0.18$ on 2019 July 17 UTC. 
We should note that the value derived from timestamps of the image data have a constant offset of several seconds due to imperfect time synchronizations of the capture software \citep{Arimatsu17}.
The best-fit Pluto $+$ Charon flux ratio is $0.261\pm 0.012$, which is slightly higher but comparable to that observed before the occultation event ($0.241$, see Section~2.2).

\subsection{Pressure evolution}
Figure~3 shows the reference radius pressure $p_{1215}$ as a function of time; our result is compared with pressure values obtained by previous studies \citep{Meza19} using the same atmosphere model profile.
The present pressure value is the lowest among those reported after 2012.
We report a diminution of pressure of $21^{+4}_{-5} \%$ ($1\sigma$ error bars) 
between 2016 (when $p_{1215} = 6.61 \pm  0.22~{\rm \mu bar}$, \citealt{Meza19}) and our measurement in 2019 
($p_{1215}  = 5.20^{+0.28}_{-0.19}~{\rm \mu bar}$, see above). 
This corresponds to a decrease rate of $7.1^{+1.2}_{-1.7} \%$ per Earth year. 
Examination of the $\chi^2$ map in Figure~2, however, shows that this drop is detected at the $2.4\sigma$ level, and thus remains marginally significant.
Previous occultation studies using the same atmospheric model profile have reported a continuous increase of the pressure
 by a factor of approximately three between 1988 and 2016 \citep{Meza19}, corresponding to an average increase rate of 
 $\sim 4\%$ per Earth year.
Taken at face value, our result thus indicates that Pluto's atmospheric pressure is decreasing nearly twice as rapidly as it increased in the previous three decades.

\section{Discussion}
The pressure evolution of Pluto's atmosphere has been due to seasonal cycles of surface volatiles, especially of ${\rm N_2}$ \citep{Hansen96, Young13, Hansen15}.
Recent exploration by NH gives a much tighter constraints on seasonal cycle models of Pluto's surface volatiles \citep{Bertrand16, Forget17, Bertrand18, Meza19}, several of which predict an ongoing (or near-future) decrease in Pluto's atmospheric pressure.  

According to a seasonal volatile transport model by \citet{Bertrand16}, the observed trend is related to the seasonal sublimation-condensation of the glacier in Sputnik Planitia.
Since the sub-solar point was close to the latitudes of northern areas ($30\degr$N - $50\degr$N)  of Sputnik Planitia in $\sim 1988-2015$, the insolation and the sublimation of ${\rm N_2}$ ice in its glacier reached maximal. 
The model predicted that the pressure reached its peak value in $\sim 2015$ and then decreased because the subsolar point moves to higher latitudes, leading to an enhanced ${\rm N_2}$ condensation rate. 
A recent model proposed by \citet{Meza19} also predicted a similar pressure drop trend after $\sim 2020$.
However, our decrease rate of pressure is larger than that predicted by the seasonal evolution models. 
For instance, the \citet{Bertrand16} model predicted that the pressure decrease rate around 2019 is $\sim 0.7-1.0\%$ (assuming a seasonal thermal inertia range of $500-1500\,{\rm J\, s^{-1/2} \,m^{-2}\, K^{-1}}$) per Earth year, approximately an order of magnitude smaller than our estimated value.
\citet{Meza19} predicted almost no change or slight upward trend in the pressure around 2019 for their assumed ${\rm N_2}$-ice albedo range ($0.72-0.73$).
These models assumed the decrease of the atmosphere mainly attributed to the condensation of ${\rm N_2}$ in Sputnik Planitia.
The present result would imply that the current ${\rm N_2}$ condensation rate at Sputnik Planitia is significantly higher than that predicted by the existing models.
Another possibility is that topographic features other than Sputnik Planitia may play an important role in a rapid ${\rm N_2}$ ice condensation. 
The present rapid decrease may imply an additional non-negligible deposit(s), especially in the hemisphere opposite to the one better observed by NH (so-called "far side" hemisphere, \citealt{Stern19}).
In any case, further continuous observations are required to obtain more detailed constraints. 

\section{Conclusion and future prospects} 
Single-chord high-speed photometric observations of the stellar occultation by Pluto on 2019 July 17 provided its latest atmospheric pressure.
Contrary to the recent increasing pressure trend revealed by previous occultation studies, 
our analysis showed the rapid (approximately $21 \%$ of the previous value) pressure drop between 2016 and 2019,
but at a 2.4$\sigma$ significance level which is marginally significant.
This decreasing trend is possibly caused by ongoing ${\rm N_2}$ condensation processes associated with the orbitally driven decline of solar insolation in Sputnik Planitia ${\rm N_2}$ ice. 
However, the observed amplitude of the pressure decrease is larger than the model predictions. 

Our single-date and single-chord
data do not bring further constraints on the seasonal evolution. 
Continuous ground-based stellar occultation observations are thus important to assess and to understand the transitional phase of Pluto's atmospheric pressure trend.
However, as already mentioned in Section~1, since Pluto is receding away from the Galactic plane, the probability of occultation events of bright stars is decreasing. 
High-sensitivity and high-cadence observations with CMOS cameras mounted on $\sim$ m class telescopes will become more important to observe occultation events of fainter stars (with magnitudes fainter than $\sim 15$, e.g., \citealt{Arimatsu19b}), which occur more frequently than those of brighter stars.
Observations with airborne telescopes such as SOFIA \citep{Person19} and with portable telescope systems for high-cadence observations (e.g., \citealt{Arimatsu17,Arimatsu19a}) will be also useful to observe rare and local occultation events.

\begin{acknowledgements}
We thank Dr. Bruno Sicardy for careful reading and providing constructive suggestions.
This research has been partly supported by Japan Society for the Promotion of Science (JSPS) Grants-in-Aid for Scientific Research (KAKENHI) Grant Numbers 15J10278, 16K17796, and 18K13606.
\end{acknowledgements}

\clearpage
\begin{figure}[ht!]
\centering
\includegraphics[width=\hsize]{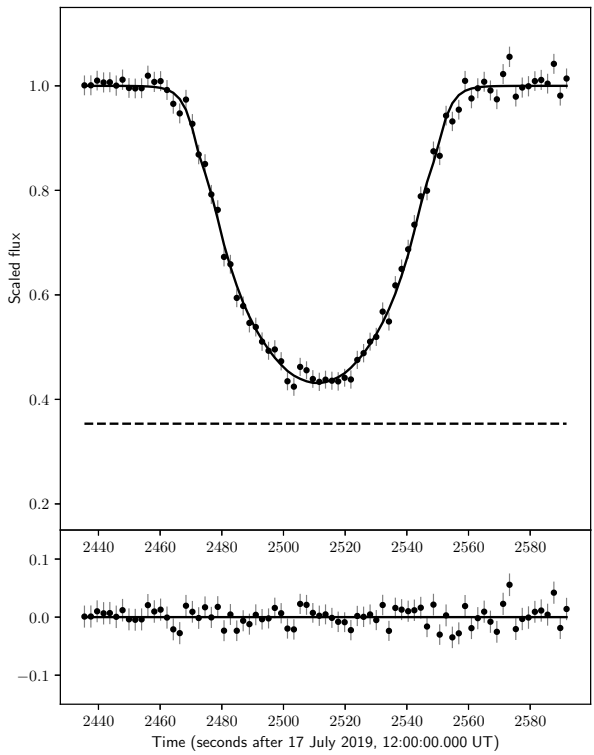}
\caption{Light curve of the occultation obtained with T60.
The data are normalized to the unocculted flux.
An error bar for each data point represents the detector readout noise, target shot noise, and sky background noise.
The sky background noise is the dominant noise source since the background level was irregularly high due to the apparent proximity of Moon during the observations (angular distance to the object corresponds to $\sim 10\deg$).
The solid curve corresponds to the best-fit synthetic light curve.
The horizontal dashed line shows the best-fit Pluto + Charon contribution to the total flux, corresponding to $0.261$.
The residuals are shown in the bottom panel.
}
\end{figure}

\clearpage
\begin{figure}[ht!]
\centering
\includegraphics[width=\hsize]{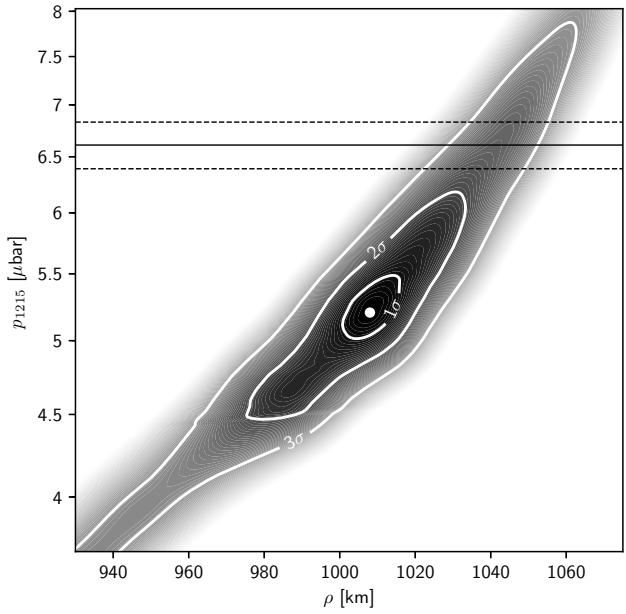}
\caption{
$\chi^2$ map for the fit to the light curve as  a function of the distance of closest approach of Haleakala Observatory to Pluto's shadow center, $\rho$, and the pressure at a radius of $1215~{\rm km}$, $p_{1215}$, which has been referred to be a reference radius by previous studies \citep{Yelle97, Sicardy16, Meza19}. 
Darker zones correspond to lower $\chi^2$ values.
The white circular point is the best fit value corresponding to the minimum $\chi^2$ value of $\chi^2_{\rm min} = 61.9$ with 73 degrees of freedom.
The inner and outer closed curves correspond to $1\sigma$ (provided by $\chi^2_{\rm min} + 1$), $2\sigma$ ($ \chi^2_{\rm min} + 4$), and $3\sigma$ ($\chi^2_{\rm min} + 9$) levels, respectively.
The solid horizontal line shows $p_{1215}$ value obtained in 2016 ($p_{1215} = 6.61 \pm  0.22~{\rm \mu bar}$, \citealt{Meza19}), and the upper and lower dashed lines correspond to its $1\sigma$ upper and lower limits, respectively.
}
\end{figure}

\clearpage
\begin{figure}[ht!]
\centering
\includegraphics[width=\hsize]{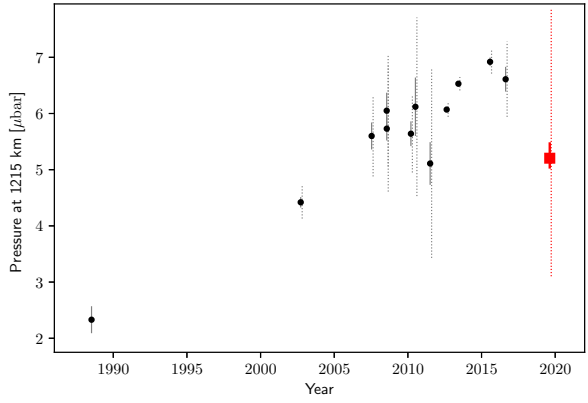}
\caption{
The atmospheric pressure of Pluto at $r = 1215~{\rm km}$, $p_{1215}$, as a function of time. A red square 
is the pressure value obtained in our present study (see Section~3), overlaid with those obtained between 1988 and 2016 using the same atmosphere model profile (black points; \citealt{Meza19}; Sicardy, private communication).
The solid and dashed lines (horizontally shifted by +0.1 years for better visibility) represent $1\sigma$ and $3\sigma$ error bars for the data points, respectively. 
Note that the $3\sigma$ error bar of the 1988 June 09 datapoint is not shown since the corresponding error value was not provided.
}
\end{figure}
\clearpage

\setlength{\tabcolsep}{2.1mm}
\renewcommand{\baselinestretch}{1.28}
\begin{table*}
\renewcommand{\arraystretch}{0.8}
\centering
\caption{Adopted parameters for the light curve fit and results}
\label{tab_param}
\begin{tabular}{llll}
\hline\hline
\multicolumn{4}{c}{Adopted parameters and physical constant} \\ 
\hline 
\hline 
\multicolumn{2}{l}{Pluto's geocentric distance} & \multicolumn{2}{l}{$4.91063 \times 10^9$~km} \\
\multicolumn{2}{l}{Pluto's shadow velocity at 12$^{\rm h}$ 42$^{\rm m}$ on 2019 July 17} & \multicolumn{2}{l}{24.18 km s$^{-1}$ (geocentric), 24.51 km s$^{-1}$ (Haleakala sta.\tablefootmark{a})} \\
\multicolumn{2}{l}{Pluto's mass and radius\tablefootmark{b}}              & 
\multicolumn{2}{l}{$GM = 8.696 \times 10^{11}$ m$^3$ s$^{-2}$, $R_P = 1187$~km} \\
\multicolumn{2}{l}{${\rm N_2}$ molecular mass}                                       & \multicolumn{2}{l}{$\mu  = 4.652 \times 10^{-26}$~kg} \\
\multicolumn{2}{l}{${\rm N_2}$ molecular refractivity\tablefootmark{c}}    & \multicolumn{2}{l}{$K = 1.091 \times 10^{-23}$} \\ 
\multicolumn{2}{l}{Boltzmann constant}      &  \multicolumn{2}{l}{$k_{\rm B}= 1.380626 \times 10^{-23}$ J K$^{-1}$} \\
\hline
\hline
%%%%%%%%%%%%%%%%%%%%%%%%%%%%%%%%%%%%%%%%%%%%%%%%%%%%%%%%%%%%%%%%%%%%%%
\multicolumn{4}{c}{Results} \\
\hline
%%%%%%%%%%%%%%%%%%%%%%%%%%%%%%%%%%%%%%%%%%%%%%%%%%%%%%%%%%%%%%%%%%%%%%
\multicolumn{4}{c}{Pressure} \\
%%%%%%%%%%%%%%%%%%%%%%%%%%%%%%%%%%%%%%%%%%%%%%%%%%%%%%%%%%%%%%%%%%%%%%
\hline
\multicolumn{2}{l}{Pressure at 1215~km, $p_{1215}$} & \multicolumn{2}{l}{$5.20^{+0.28}_{-0.19}~{\rm \mu bar}$}  \\
\multicolumn{2}{l}{Surface pressure, $p_{\rm surf}$} & \multicolumn{2}{l}{$9.56^{+0.52}_{-0.34}~{\rm \mu bar}$}  \\
\hline
%%%%%%%%%%%%%%%%%%%%%%%%%%%%%%%%%%%%%%%%%%%%%%%%%%%%%%%%%%%%%%%%%%%%%%
\multicolumn{4}{c}{Astrometry} \\
%%%%%%%%%%%%%%%%%%%%%%%%%%%%%%%%%%%%%%%%%%%%%%%%%%%%%%%%%%%%%%%%%%%%%%
\hline
\multicolumn{2}{l}{Closest approach to Pluto's shadow center}         &  \multicolumn{2}{l}{$1008.0^{+7.8}_{-7.2}$~km} \\
\multicolumn{2}{l}{Time of closest approach to shadow center (UT)\tablefootmark{c}} & \multicolumn{2}{l}{12$^{\rm h}$ 41$^{\rm m}$ $52.02\pm 0.18$$^{\rm s}$ on 2019 July 17} \\
\hline
\end{tabular}
~  \\
\raggedright
\tablefoottext{a}{For the Earth rotation correction, the WG84 shape model is used.}
\tablefoottext{b}{\cite{Stern15}, where $G$ is the constant of gravitation.}
\tablefoottext{c}{\cite{was30}.} 
\tablefoottext{d}{A possible systematic bias is several seconds. (see Section 3.1)}
\end{table*} 
\renewcommand{\baselinestretch}{1.6}

\end{document}